\documentclass[aip,jcp,amsmath,amssymb, reprint,xelatex]{revtex4-1}
\usepackage[english]{babel}
\usepackage{graphicx}
\usepackage{array}
\usepackage{bm}
\usepackage[utf8]{inputenc}
\usepackage[T1]{fontenc}

\begin{document}
\title{Ab-initio Structure and Thermodynamics of the RPBE-D3 Water/Vapor Interface by Neural-Network Molecular Dynamics}
\author{Oliver Wohlfahrt}
\affiliation{University of Vienna, Faculty of Physics, Boltzmanngasse 5, A-1090, Vienna, Austria}
\author{Christoph Dellago}
\email{christoph.dellago@univie.ac.at}
\affiliation{University of Vienna, Faculty of Physics, Boltzmanngasse 5, A-1090, Vienna, Austria}
\author{Marcello Sega}
\email{m.sega@fz-juelich.de}
\affiliation{University of Vienna, Faculty of Physics, Boltzmanngasse 5, A-1090, Vienna, Austria}
\affiliation{Forschungszentrum Jülich GmbH, Helmholtz Institute Erlangen-Nürnberg for Renewable Energy (IEK-11), Fürther Str. 248, D-90429, Nürnberg, Germany}
\date{\today}

\begin{abstract}
	Aided by a neural network representation of the density functional theory (DFT) potential energy landscape of water in the RPBE approximation corrected for dispersion, we calculate several structural and thermodynamic properties of its liquid/vapor interface. The neural network speed allows us to bridge the size and time scale gaps required to sample the properties of water along its liquid/vapor coexistence line with unprecedented precision.
\end{abstract}

\maketitle

\section{Introduction}
The strong, directed network of hydrogen bonds\cite{kumar_hydrogen_2007} that confers water
its rich phase diagram, and its numerous anomalous properties\cite{errington_relationship_2001}, is also
responsible for the peculiar structure of its liquid/vapor interface\cite{sovago_hydrogen_2009}.
Thanks to its hydrogen bond network, water is one of the most cohesive liquids known, with exceptionally high surface tension. 

The anomalous surface tension is, by far, not the only surprising property of water interfaces. Unlike simple liquids, water undergoes substantial structural changes in the interface's first molecular layer. In its preferential configuration, a surface molecule at the liquid/vapor interface reorients its dipole vector along the liquid surface, with one OH bond and one lone pair both directed towards the vapor phase\cite{morita_theoretical_2000,scatena_water_2001,kuhne_new_2011}. The rearrangement of water molecules at the interface with other phases, with the corresponding entropic change, lies at the very heart of phenomena like the hydrophobic effect\cite{chandler_interfaces_2005}, one of the main actors of self-organization in living organisms, and of many physicochemical based applications, from detergents to novel materials based on microemulsions.

Understanding the relationship between the microscopic structure
of the water/vapor interface and its thermodynamic properties is
then key to obtaining better control over a vast array of processes.
Due to its molecular-scale extension, interfacial water can be
investigated experimentally by a small set of techniques, most
importantly, X-ray reflectivity\cite{braslau_surface_1985} and
sum-frequency generation spectroscopy\cite{shen_surface_1989}. In
this sense, computer simulations techniques, providing direct access
to the molecular configurations, are a unique asset. However, the
accuracy of the calculations and their computational cost are two
significant challenges for the simulation of water in general and
its interfaces in particular. Ideally, one would aim at an accurate
parameter-free description. While much progress has been made to
achieve better ab-initio simulations of liquid
water\cite{gillan_perspective_2016,chen_ab_2017}, the computational
resources required to simulate liquid interfaces using state-of-the-art
methods are still daunting. Artificial neural networks have proven
to be a valuable tool\cite{bartok2017machine} to reproduce the features of the potential
energy landscapes of water\cite{morawietz_how_2016}. Using an artificial neural network to
reproduce the forces acting between nuclei as computed with an
ab-initio method of choice can decrease the computational cost of
the problem by several orders of magnitude\cite{behler_generalized_2007}. However, the neural
network-predicted forces associated with a specific configuration
can be unreliable if the network has not encountered sufficiently
similar configurations during its training phase.

Here, we extend previously parameterized neural networks\cite{morawietz_how_2016,morawietz_hdnnp_2019} by including explicitly interfacial configurations and use the optimized potentials to study the phase diagram of water along its liquid/vapor coexistence line. The need for relatively large systems is particularly pressing for interfacial systems as they suffer more than bulk systems from finite-size effects\cite{gelfand_finite-size_1990}. The computational advantage warranted by the neural network allowed us to simulate for many nanoseconds simulation boxes containing 1024 water molecules, based on the RPBE generalized gradient approximation for the exchange-correlation functional\cite{hammer_improved_1999} supplemented by Grimme's D3 dispersion corrections\cite{grimme_consistent_2010}. 

\begin{figure}
	\includegraphics[width=\columnwidth]{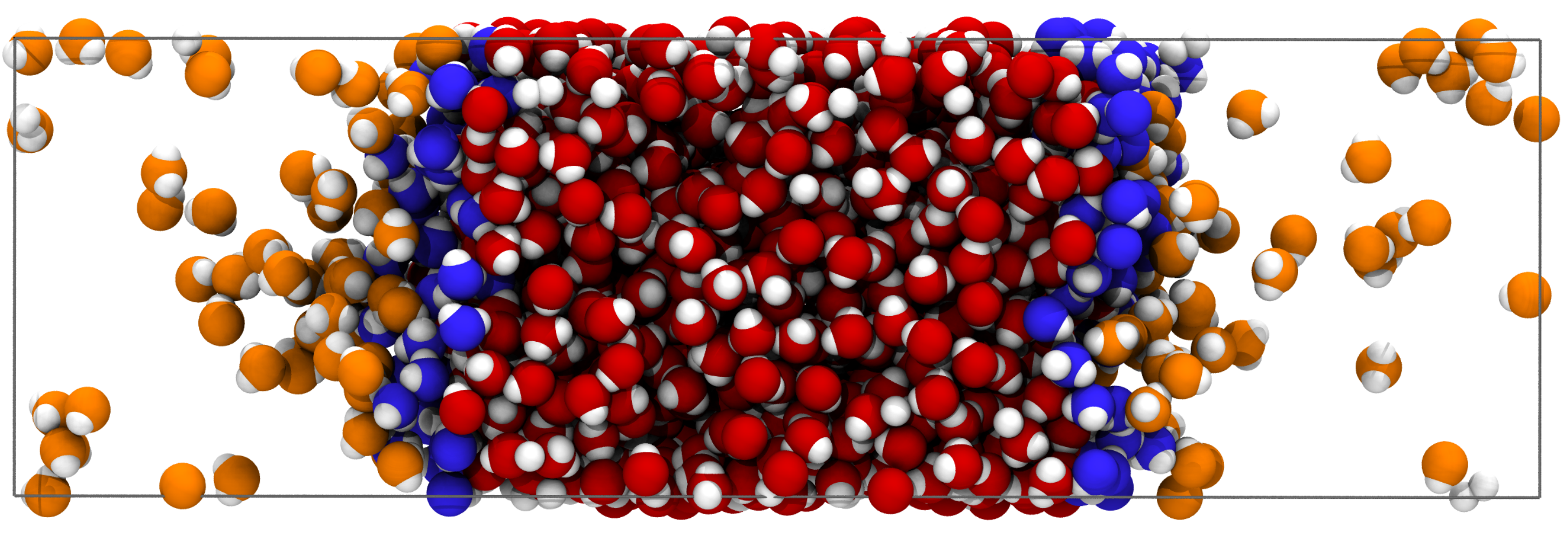}
\caption{\label{fig:snaps}
	Simulation snapshot of the simulation box of 1024 water molecules at 550~K. The oxygen atoms are colored in red (liquid phase), blue (interfacial layer) and orange (vapor phase).
}
\end{figure}

\section{Extending the Neural Network Training Set\label{sec:extension}}
As a first step, we simulated a water/vapor interface using the neural network for the RPBE exchange-correlation functional with Grimme's D3 corrections presented in Ref~\onlinecite{morawietz_how_2016}. This network was trained using configurations taken from the liquid, solid, and liquid/solid coexistence phases\cite{morawietz_hdnnp_2019}. As expected, the network could not recognize a relatively high fraction of configurations at the liquid/vapor interface, resulting in unphysically high vapor densities and large interfacial widths. A preliminary test with the SPC/F\cite{toukan_molecular-dynamics_1985} empirical model showed that using a training dataset of about 400 frames generated from equilibrium trajectories of a liquid/vapor interface (216 molecules) yields excellent convergence and physically sound results. 
The density in the liquid phase obtained by the neural network differred from explicit SPC/F simulation from a minimum of 0.6$\%$ at 300 K, up to a maximum of 3$\%$ at 460 K, while the surface tension differences were always below 1 mN/m. 

We used these 400 frames to augment the dataset presented in Ref~\onlinecite{morawietz_how_2016}, taking care of  perturbing the nuclear coordinates (by a uniform distribution up to $\pm$ 0.003 nm) to enhance the sampling of the neighborhood of the free energy minima. The DFT energies and forces at the RPBE level were calculated for the whole training dataset using FHI-AIMS\cite{blum_ab_2009} and D3 corrections were added. We then used this initial extension to train the neural network and generated a trajectory with a temperature ramp from 300 to 600K over 1 ns of integration. Next, we took about 500 (unperturbed) frames from this trajectory to extend the training set further, performing in this way a self-consistent refinement step. We report additional methodological details in Sec.\ref{sec:methods}.

Here, we would like to stress that without the extension of the training dataset no meaningful simulation of the liquid/vapour equilibrium could be performed. For example, the interface would become unstable already at about 550~K (to be compared with a critical point of 632~K), and unphysical configurations in the first liquid layer, with the majority of molecular dipoles pointing towards the gas phase, instead of being parallel to the interface. Note that the new training set with interfacial configurations includes also the bulk configurations used in Ref.~\onlinecite{morawietz_how_2016}, and the thermodynamic properties of the liquid state are compatible with those reported there, at comparable thermodynamic points.

Simulating water using the SPC/F empirical potential allowed us also to test the effect of using a cutoff in the neural network, as opposed to the inclusion of long range forces with mesh Ewald methods. We performed additional simulations with the SPC/F model at 300 K, employing the same cutoff used for the neural network (0.635 nm), which yielded a similar (1$\%$ difference) density of the liquid phase, but a remarkably different surface tension, with a difference of about 9 mN/m. This confirms, albeit indirectly, that the neural network encodes the effects of long-range forces, through the local arrangement of atoms and their forces, even if a simple cutoff is used. With the present simulations it is not possible to tell, however, to which extent the inaccuracy of the results obtained with the neural network can be ascribed to the lack of explicit information about further neighbors, by the choice of symmetry functions, or by intrinsic limitations of the neural network itself.

\section{Results\label{sec:results}}
We ran molecular dynamics simulations in the canonical ensemble using the neural network potential for systems of 1024 water molecules in slab configurations within simulation boxes of size $3\times3\times10$~nm under periodic boundary conditions, at 15 different temperature values ranging from 300 to 620~K. Each simulation started from a pre-equilibrated configuration of the SPC/F model at the corresponding temperature, using a timestep of 0.5 fs. After 100 ps, we observed no drift in the potential energy and in the density profile of each trajectory, and we deemed to have reached equilibrium. We saved configurations to disk during the subsequent 1 ns of trajectory at intervals of 1 ps for further analysis. Values of energy and pressure were dumped every 10 fs.

\begin{figure}
    \includegraphics[width=\columnwidth]{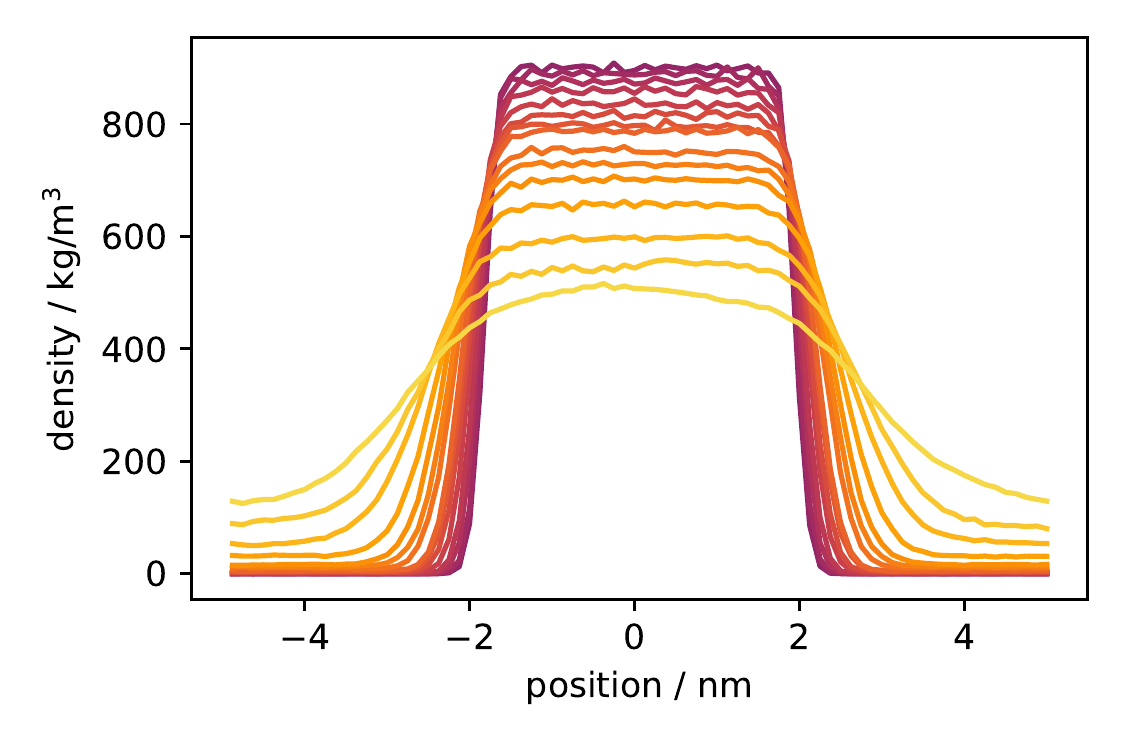}
    \caption{\label{fig:profiles}Mass density profiles across the whole simulation box, for the whole range of temperature investigated (see Tab.\ref{tab:coexistence}) from 300~K (highest density in the liquid phase) to 620~K (lowest density in the liquid phase).
    }
\end{figure}

\begin{figure}
    \includegraphics[width=\columnwidth]{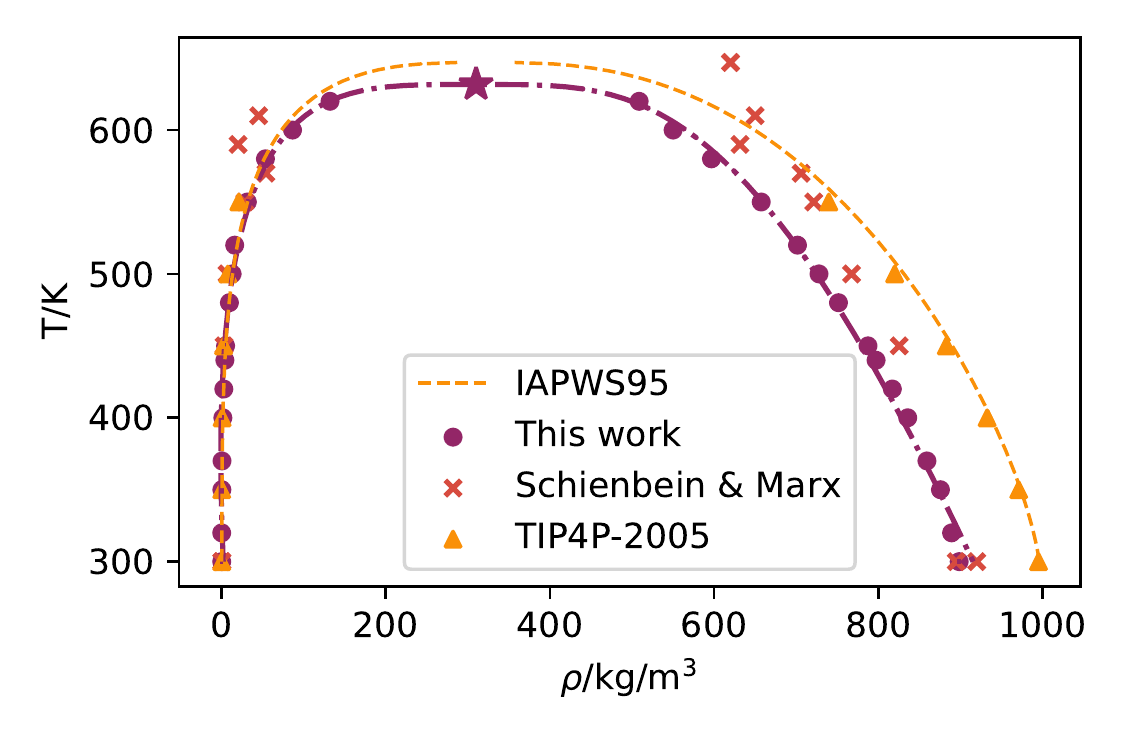}
    \caption{\label{fig:coexistence}Water liquid/vapor coexistence from simulations using the RPBE-D3 approximation (circles, this work; crosses, Ref.~\onlinecite{schienbein_liquidvapor_2018}) and using the TIP4-2005 empirical potential (triangles,  from Ref.~\onlinecite{sega_long-range_2017}). Best fit of the data from the present work to Eq.(\ref{eq:wilding}) (dot-dashed line), corresponding estimate of the critical point (star) and the IAPWS95 equation of state (dashed line). }
\end{figure}

From the mass density profiles reported in Fig.~\ref{fig:profiles}, one can appreciate the progressive broadening of the slab width, accompanied by the density increase in the vapor phase and corresponding decrease in the liquid one. In Fig.~\ref{fig:coexistence} we report the density values of the 1 nm-wide regions located in the middle of the liquid and vapor slab.
To extrapolate the critical point location, we performed the best fit of the simulation results to the expression proposed by Wilding\cite{wilding_critical-point_1995},
\begin{equation}
\rho(T) = \rho_c + a \left| T-T_c \right| \pm b  (T-T_c)^\beta,
\label{eq:wilding}
\end{equation}
where $\rho_c$ and $T_c$ are the critical density and temperature, and the sign plus and minus applies to the liquid and vapor branch, respectively. We used the scaling exponent $\beta$ as a fitting parameter, in addition to $a$ and $b$, obtaining as best fit estimates $\rho_c=310\pm3$~kg/m$^3$, $T_c=632\pm2$~K, $\beta=0.27\pm0.01$, $a=0.45\pm0.01$ and $b=94\pm5$. 
 
In Fig.~\ref{fig:coexistence} we report also the IAPWS95 curve\cite{wagner_iapws_2002}, which matches the experimental data, and the points from the only other estimate of the coexistence line of RPBE-D3 water we are aware of, taken from the work of Schienbein and Marx\cite{schienbein_liquidvapor_2018}. Notice the steeper trend of Schienbein and Marx's data, obtained via ab-initio Gibbs ensemble Monte Carlo\cite{panagiotopoulos_direct_1987,mcgrath_simulating_2006} of 128 water molecules, which is arguably a finite-size effect yielding an effective critical temperature $T_c(L)$ that shifts toward higher values with decreasing system (linear) size $L$ as\cite{mon_finite_1992}
\begin{equation}
T_c(L) - T_c\simeq \frac{1}{L^{1/\nu}},
\end{equation}
where $\nu$ is the critical exponent governing the scaling of the correlation length. 

\begin{figure}
    \includegraphics[width=\columnwidth]{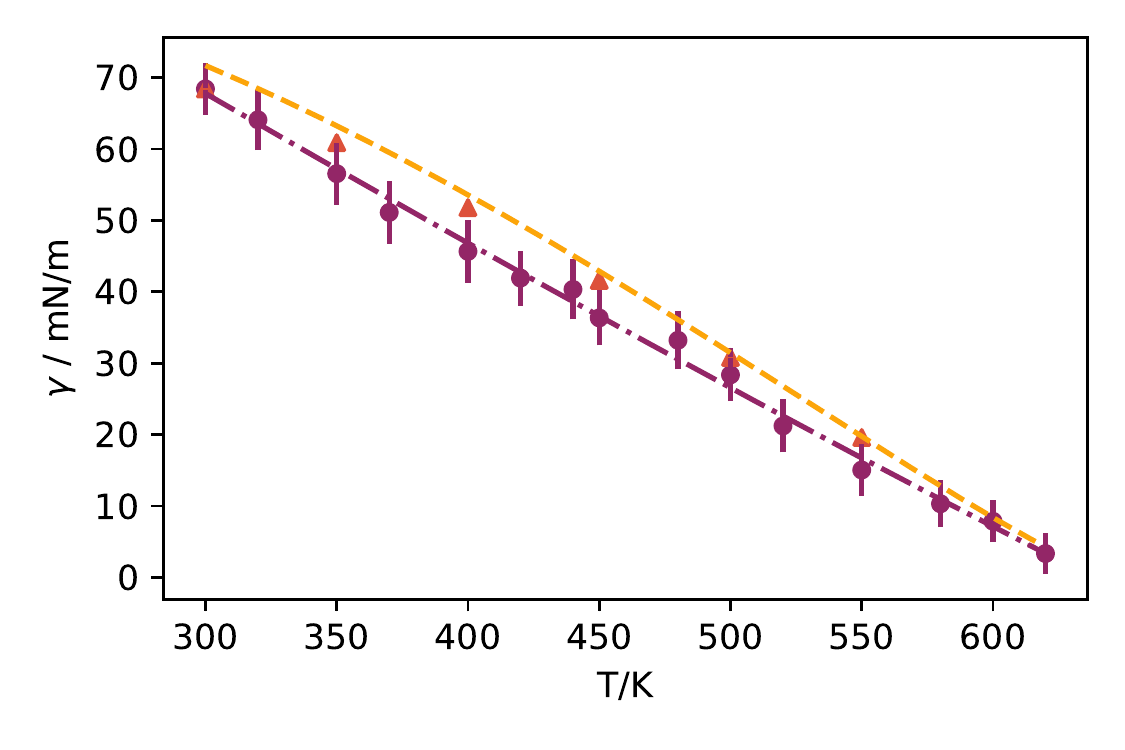}
    \caption{\label{fig:surftens}Surface tension of RPBE-D3 as computed in this work (circles) and of TIP4P-2005  (triangles, taken from Ref.~\onlinecite{sega_long-range_2017}). The dot-dashed line is the result of a best fit to Eq.~(1) from Ref.~\onlinecite{vargaftik_international_1983}. The dashed line is the best-fit to experimental data reported in Ref.~\onlinecite{vargaftik_international_1983}.}
\end{figure}

From the average values of the pressure tensor elements, $p_{ij}$, it is straightforward to compute the surface tension $\gamma$ using the mechanical route, as
\begin{equation}
\gamma = \frac{L_z}{2}\left[ p_{zz} - \left(p_{xx}+p_	{yy}\right)/2 \right],
\end{equation}
where $z$ identifies the direction normal to the macroscopic interface. In Fig.~\ref{fig:surftens} we report the surface tension as a function of temperature, the result of the best fit to Eq.(1) from Ref.~\onlinecite{vargaftik_international_1983}, and we compare them with the interpolated experimental values. Even though the present results do not match the experimental curve as well as the best empirical potential models like TIP4P-2005\cite{abascal_general_2005}, the agreement is still very good, and superior to several other mainstream empirical models\cite{sega_long-range_2017}.

\begin{table}\begin{tabular}{lrrr}
\hline\hline  
$T$ / K & $\rho_l$ / kg/m$^3$ & $\rho_g$ / kg/m$^3$ & $\gamma$ / mN/m \\
\hline
300& 899(1) & 0.03(1)&68(2)\\
320& 889(1) & 0.02(1)&64(2)\\
350& 876(1) & 0.20(2)&57(2)\\
370& 859(1) & 0.37(5)&51(2)\\
400& 836(1) & 1.41(5)&46(2)\\
420& 817(1) & 2.57(8)&42(2)\\
440& 797(1) & 3.9(2) &40(2)\\
450& 788(1) & 4.7(1) &36(2)\\
480& 752(1) & 9.4(3) &33(2)\\
500& 728(1) & 12.7(2)&28(2)\\
520& 702(1) & 15.8(2)&21(2)\\
550& 657(1) & 31.2(3)&15(2)\\
580& 597(1) & 53.3(6)&10(2)\\
600& 550(1) & 86(1)  &8(2) \\
620& 509(1) & 132(2) &3(2) \\
\hline
632(2)&\multicolumn{2}{c}{310(2)} &\\
\hline\hline
\end{tabular}
\caption{Liquid ($\rho_l$) and vapor ($\rho_g$) densities and surface tension ($\gamma$) as a function of the 
	temperature. The estimated critical point is reported in the last line. Values in parentheses represent one standard deviation in the least significant digits.
\label{tab:coexistence}}
\end{table}

Density and surface tension values along the coexistence line are all reported in Tab.\ref{tab:coexistence}, together with the estimated critical point.

With plenty of configurations at hand, it is now possible to investigate, in a statistically meaningful way, how water's structural features are dependent on whether the molecules are located right at the interface, or below it. Molecular dynamics simulations with empirical potentials show that, at the liquid/vapor interface, water exhibits a much weaker order than van-der-Waals liquids, and many structural and dynamical properties reach their bulk value roughly after the second interfacial layer. In our investigation, we witnessed the same behavior, and here we only concentrate on the properties of the first molecular layer, detected on a per-frame basis using the Pytim analysis package\cite{sega_pytim_2018} as described in Sec.\ref{sec:methods}. For brevity, we will refer to ``bulk properties'' extracted from the trajectories presented in this work, as those properties computed from the molecules beyond the third surface layer, where the observables already converged to position-independent values.

We calculated also the bond length and angle distributions in the first and subsequent layers, and found that the molecules in the first layer are less stretched, with a difference, at 300 K, of 0.3~pm and 0.3~deg for the bond length and angle, respectively. These differences become less pronounced when the temperature is raised.

\begin{figure}
    \includegraphics[width=\columnwidth]{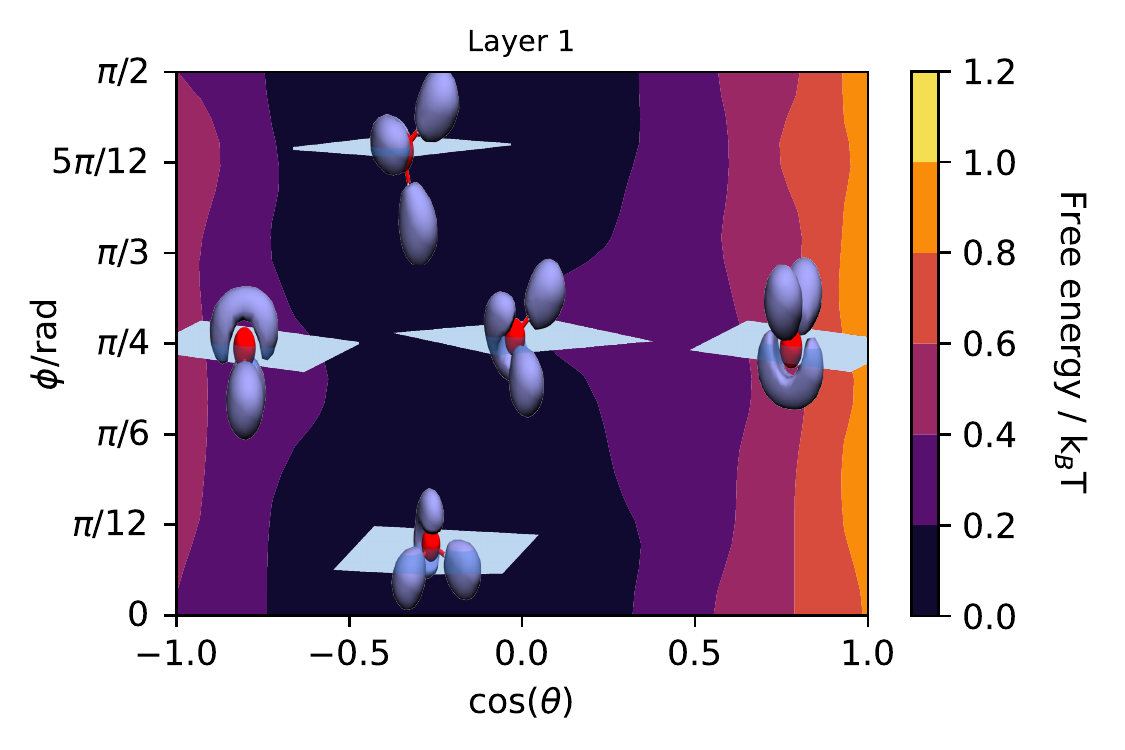}
    \caption{\label{fig:orientation} Free energy map at 300~K of the orientation of water molecules in the first molecular layer (interpolated using cubic splines). The overlaid water molecules, represented using the position of the nuclei and their electronic charge density (85\% isodensity surface) show (approximately) their orientation in the physical space with respect to the macroscopic surface plane for some selected pairs ($\phi$,$\cos\theta$).
    }
\end{figure}

Next, we characterize the orientation of the water molecules at the surface. To this purpose, we employ two angles. The first one, $\theta $, is the angle that the molecular axis vector (pointing from the oxygen atom to the midpoint between the hydrogen ones) is making with the macroscopic surface normal (pointing from the liquid to the vapor phase). The second one, $\phi$, is the angle that the molecule would need to rotate along its molecular axis to have the H-H vector aligned parallel to the surface plane\cite{morita_theoretical_2000,jedlovszky_full_2004}. In Fig.~\ref{fig:orientation}, we report the free energy of molecules in the first layer as a function of $\cos \theta$ and $\phi$, as for a randomly distributed molecular orientation, those histograms would be homogeneous.

The free energy plot shows a prevalence of molecular orientations (the minima) when the dipole moment is aligned parallel to the surface plane, or pointing slightly below it ($\cos \theta \simeq -0.25$) and when the OH bonds either laying in the surface plane ($\phi\simeq 0$) or pointing out of it ($\phi\simeq \pi/2$). This result agrees qualitatively with previous experimental findings and simulations with empirical potentials. The orientations with the dipole moment pointing toward the vapor ($\cos\theta=1$) or the liquid ($\cos\theta=-1$) are less likely to be found than the parallel orientation, although all orientations are accessible within an energy of $k_\mathrm{B}T$, where $k_\mathrm{B}$ is the Boltzmann constant. When the temperature increases the free energy map becomes gradually flatter and although the preferential orientation of the dipole vector is always parallel to the surface, above 550K all orientations are accessible within an energy of 0.2 $k_\mathrm{B}T$ (see Supplementary Material).

\begin{figure}
    \includegraphics[width=\columnwidth]{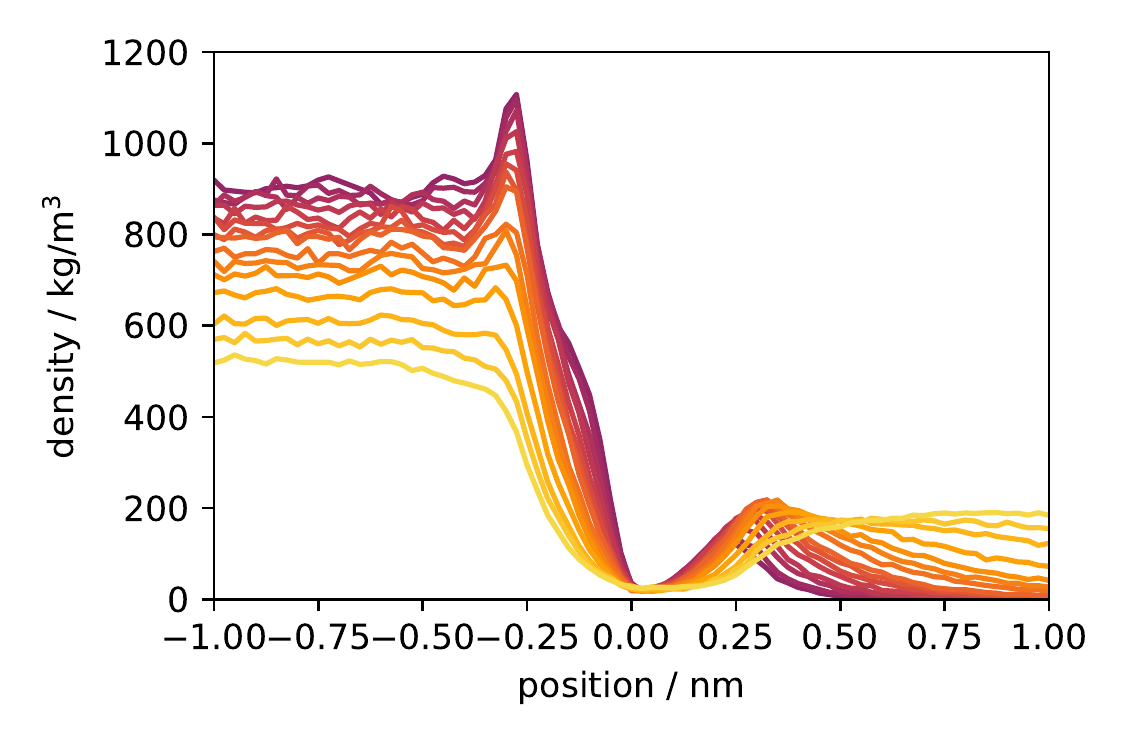}
    \includegraphics[width=\columnwidth]{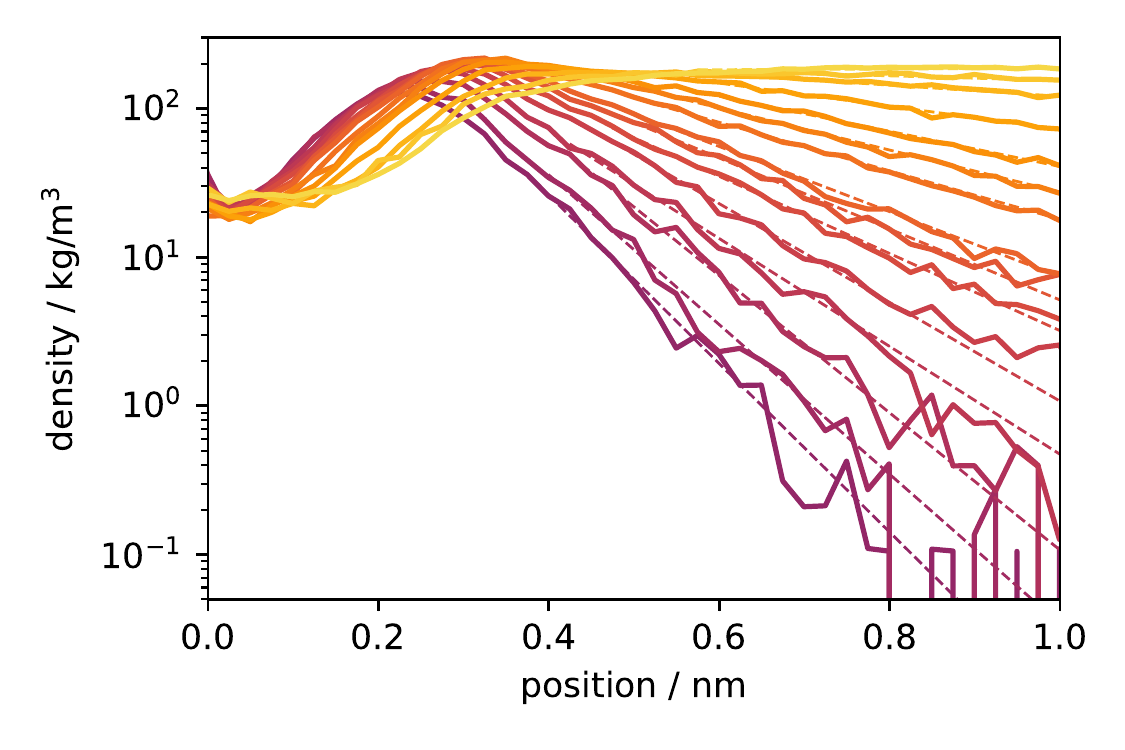}
    \caption{\label{fig:intrinsic}Upper panel: intrinsic mass density profiles
of water molecules next to the interface for the whole range of
temperature investigated (see Tab.\ref{tab:coexistence}) from
300~K (highest density in the liquid phase) to 620~K (lowest
density in the liquid phase). The vapor phase is located by
convention at positions on the right (positive), relative to 
the surface. The delta-like contribution at the origin is not shown.
Lower panel: detail of the vapor side close to the interface, in
semilogarithmic scale. Thin dashed lines are the result of best
fits to exponential decays to the bulk vapor density.}
\end{figure}

Having access to the set of surface molecules, it is possible to build a local reference frame $(x,y,\zeta(x,y))$ on the corrugated interface and use it to compute the intrinsic density profile\cite{chacon_intrinsic_2003}, as reported in Fig.~\ref{fig:intrinsic}. The usual density profile expresses the correlation between molecular positions in the system and the location of its center of mass. In contrast, the intrinsic density profile, $\rho_I(z')$, represents the correlation between molecular positions and the local position of the interface,
\begin{equation}
\rho_I(z') = \frac{m}{A}\left\langle \sum_i^N \delta(z' - z_i + \zeta(x_i,y_i) ) \right\rangle,
\end{equation}
where $A$ is the simulation box cross-sectional area, $N$ is the
number of molecules, $m$ their mass, and angular brackets stand for
the canonical average. The molecules in the surface layer
are located at $z'=0$, giving rise to a delta peak, which is not
shown in Fig.~\ref{fig:intrinsic}. By convention, positive values
of $z'$ are located in the vapor phase. At relatively low temperature,
where the vapor density is small, and the liquid's cohesive strength
is the highest, the structure of the local packing emerges with a
clear peak and a small shoulder, similar to the results from
simulations with empirical potentials. With increasing temperature,
the local structure at the interface becomes less pronounced and
almost disappears at 620 K. On the vapor side, one observes a similar
behavior, with a peak of the vapor density next to the interface that
vanishes at high temperature. One can expect condensing vapor at the
interface because the molecules in the vapor phase feel both the
attractive dispersion forces and residual dipolar interactions.  Upon
a closer look at the vapor phase region in a semi-logarithmic scale
(Fig.~\ref{fig:intrinsic}, lower panel), we notice that the density,
beyond the local maximum, decays exponentially toward the bulk vapor
density like $\exp(-z/\xi)$, as it is expected for a dilute vapor
next to a weakly attractive surface (the liquid phase) in the mean
field approximation\cite{henderson_fundamentals_2009,fisk_structure_1969}.
The closer to the critical temperature, the more the profile tends
toward homogeneity (with vapor density approaching the liquid one) except for 
an excluded volume region right at the interface that resembles the
square-well one would expect from a hard-sphere.

\section{Methods\label{sec:methods}}
The neural network potential has been set up using the approach by Behler and Parrinello\cite{behler_generalized_2007}, using the implementation \texttt{n2p2} of Singraber and coworkers\cite{singraber_library-based_2019}. We used the same selection of symmetry functions and cutoff values reported in Ref.~\onlinecite{morawietz_how_2016}. For the sake of consistency and simplicity, this network is not trained to reproduce charge distributions (although is in principle possible). Hence, no electrostatic quantities can be directly computed from the present simulation results. The network was able to reproduce the forces acting on atoms within 1.5 meV/\AA{} (1 standard deviation, $\simeq68\%$), 2.2 meV/\AA{} (2 standard deviations, $\simeq95\%$) and 4 meV/\AA{} (3 standard deviations, $\simeq99.7\%$), with a typical rms error of 3\% (see Supplementary Material).
DFT calculations were performed using the FHI-AIMS package\cite{blum_ab_2009}, with second tier level of basis functions for oxygen and hydrogen atoms, reproducing without any significant difference the forces and energies of the old dataset of Ref~\onlinecite{morawietz_how_2016}.
Neural network molecular dynamics simulations were performed using LAMMPS (\texttt{http://lammps.sandia.gov})\cite{plimpton_fast_1995}, linked to the neural network library of Singraber and coworkers (\texttt{https://github.com/CompPhysVienna/n2p2})\cite{singraber_library-based_2019}. Each system consisted of 1024 water molecules simulated in the canonical ensemble using the Nos\'e--Hoover thermostat (damping constant 0.5 ps) and an integration timestep of 0.5~fs. The center of mass was prevented from drifting by subtracting the center of mass velocity at every step, and the liquid slab was kept in this way in the middle of the simulation box along the surface normal direction.
The surface layer was determined using the MDAnalysis\cite{michaud-agrawal_mdanalysis_2011}-based pytim package (\texttt{https://github.com/Marcello-Sega/pytim})\cite{sega_pytim_2018} via a combination of the ITIM\cite{partay_new_2008} method (probe sphere radius 2 \AA) and DBSCAN\cite{ester1996density}-based prefiltering of the vapor phase\cite{sega_phase_2017}, using the automatic determination of the density threshold for the clustering procedure.\\

\section{Conclusions\label{sec:conclusions}}
We have performed what is arguably the most accurate calculation, to-date, of the liquid/vapor coexistence of water described by the RPBE exchange-correlation functional, supplemented by dispersion corrections. This result was possible thanks to the use of a neural network-based fit of the DFT potential energy surface. We reported the coexistence curve of the system, estimated the critical temperature of the model ($T_c=632\pm2$~K), the surface tension curve as a function of temperature, and two order parameters, namely, the density profile and  the orientation of water molecules in the surface layer. While the most refined empirical potential models are still superior in describing some aspects of the thermodynamics of water at interfaces, ab-initio calculations are becoming increasingly more accurate, albeit still very expensive computationally. Neural network-based approaches like the present one, alone or by exploiting promotion to the DFT level of choice\cite{cheng_ab_2019}, open up the possibility to explore with superior statistical accuracy systems that were, until now, almost exclusively in the realm of empirical potential-based simulations.

\section*{Supplementary Material}
See supplementary material for the neural network training forces histogram and the orientation free energy maps.

\section*{Acknowledgments}
The computational results presented have been achieved using the Vienna Scientific Cluster (VSC).

We thank Andreas Singraber for providing help with the \texttt{n2p2} package.

\section*{Data Availability}
The datasets of the neural network training configurations with forces, the optimal weights, as well as the input files are openly available in Zenodo at \texttt{http://doi.org/10.5281/zenodo.3944892}, reference number 3944892.


%

\end{document}